\documentclass[aps,prl,twocolumn,showpacs,preprintnumbers,superscriptaddress,amsmath,amssymb]{revtex4}
\usepackage{amsmath, amsthm,textcomp }
\usepackage{graphicx}
\usepackage{dcolumn}
\usepackage{bm}
\usepackage{amssymb}
\usepackage{natbib}
\usepackage{array}
\usepackage{multirow}

\makeatletter

\newcommand{\be}{\begin{equation}}
\newcommand{\ee}{\end{equation}}
\newcommand{\ba}{\begin{array}}
\newcommand{\ea}{\end{array}}
\newcommand{\bn}{\begin{eqnarray}}
\newcommand{\en}{\end{eqnarray}}
\newcommand{\bt}{\begin{tabular}}
\newcommand{\et}{\end{tabular}}
\newcommand{\bml}{\begin{mathletters}}
\newcommand{\eml}{\end{mathletters}}
\newcommand{\bc}{\begin{center}}
\newcommand{\ec}{\end{center}}
\newcommand{\bi}{\begin{itemize}}
\newcommand{\ei}{\end{itemize}}

\newcommand{\abi}{{\it ab initio}}
\newcommand{\rmd}{\mbox{d}}

\begin{document}

\title{Ab Initio Derivation of Model Energy Density Functionals}

\author{Jacek Dobaczewski}
\affiliation{Department of Physics, University of York, Heslington, York YO10 5DD, United Kingdom}
\affiliation{Department of Physics, P.O. Box 35 (YFL),
University of Jyv\"askyl\"a, FI-40014  Jyv\"askyl\"a, Finland}
\affiliation{Institute of Theoretical Physics, Faculty of Physics, University of Warsaw,
Pasteura 5, PL-02-093 Warsaw, Poland}
\affiliation{Helsinki Institute of Physics, P.O. Box 64, FI-00014 Helsinki, Finland}

\begin{abstract}
I propose a simple and manageable method that allows for deriving
coupling constants of model energy density functionals (EDFs) directly from
{\abi} calculations performed for finite fermion systems. A
proof-of-principle application allows for linking properties of
finite nuclei, determined by using the nuclear nonlocal Gogny
functional, to the coupling constants of the quasilocal Skyrme
functional. The method does not rely on properties of infinite
fermion systems but on the {\abi} calculations in finite systems.
It also allows for quantifying merits of different model
EDFs in describing the {\abi} results.
\end{abstract}

\maketitle

Approaches based on the density functional theory (DFT) provide us
with very efficient and useful tools to describe properties of
many-fermion systems like molecules, solids, or
nuclei~\cite{[Par89],[Dre90],[Pet91],[Tsu14]}. The range of possible various
applications and implementations is extremely wide. In
electronic systems, there exist numerous methods and techniques of
linking the density functionals to the underlying Coulomb
interaction, but in nuclei such links are much more difficult to
explore, primarily because of the fact that nucleon-nucleon
interactions are less obviously definable.

Only fairly recently, families of interactions based on chiral
effective field
theory~\cite{[Bog03],[Epe06],[Epe09],[Bog10],[Mac11],[Lyn14],[Eks15]}
have become a gold standard for {\abi} calculations of nuclear
properties~\cite{[Bar13],[Som13],[Her13],[Pin13],[Hag14],[Her14],[Som14],[Hol14]}.
Based on this developments, there were already
many attempts to link the nuclear DFT to these {\abi}
approaches~\cite{[Fur08a],[Dru10],[Kai10],[Sto10],[Hol11],[Bog11],[Gan11],[Kai13]}.
Up to now, most of them referred directly (or indirectly, through the
so-called density-matrix expansion (DME)~\cite{[Neg72]}) to
infinite-nuclear-matter properties. In this paper, I propose a
generic method that directly links the {\abi} approaches to the DFT
in finite fermion systems.

A classic formulation of the DFT relies on a variational approach
to the many-body problem, whereby it assumed that we are able to
perform an exact variation of the average energy
$\delta{E}=\delta\langle\Psi|\hat{H}|\Psi\rangle=0$, which gives us
the exact ground-state energy $E_0$ and the exact ground-state
wave-function $|\Psi_0\rangle$. By replacing the full variation with a
two-stage variation, the DFT then appears in a very natural way.

The first-stage variation is performed under the constraint
that the one-body local density of the system,
$\rho(\bm{r})=\langle\Psi|a^+_{\bm{r}}a_{\bm{r}}|\Psi\rangle$, is
fixed to a given density profile. Performing such a variation for all
density profiles, one obtains the exact energy density functional (EDF)
$E\left[\rho\right]$.
In principle, such a constrained variation can be realized by an
unconstrained variation performed for the system placed in an external
one-body local potential $-U(\bm{r})$, that is,
\be\label{eq:00}
\delta{E'}=\delta\langle\Psi|\hat{H}-\hat{U}|\Psi\rangle=
\delta\left[\langle\Psi|\hat{H}|\Psi\rangle-
\!\!\int\!\!\rmd\bm{r}U(\bm{r})\rho(\bm{r})\right]=0,
\ee
whereupon the external potential acquires a role of the Lagrange
multiplier. In a sense, the constrained variation corresponds to
probing the system with an external one-body field. By inverting the
obtained relation $\rho\left[U\right]$ and inserting it into the
functional $E\left[U\right]$, one can, again in principle, obtain the
final exact EDF $E\left[\rho\right]$. These classic arguments have the
same structure as the effective-action approach, see, e.g.,
Refs.~\cite{[Dru10],[Fur12],[Kem13]}.

The second-stage variation, with respect to the density,
$\delta_{\rho(\bm{r})}E\left[\rho\right]=0$, obviously then gives the
exact ground-state energy $E_0$ and the exact ground-state local
one-body density
$\rho_0(\bm{r})=\langle\Psi_0|a^+_{\bm{r}}a_{\bm{r}}|\Psi_0\rangle$.
It is also obvious that the above argumentation can be repeated
{\it mutatis mutandis} for a functional of a one-body non-local
density $\rho(\bm{r},\bm{r}')=\langle\Psi|a^+_{\bm{r}'}a_{\bm{r}}|\Psi\rangle$,
which is the formulation we are concerned with below.

Certainly, such an idealistic derivation of the exact EDF would defy
its purpose: had we been able to perform the exact variational
calculations for all one-body potentials $-U(\bm{r})$, we would have
probably not need DFT at all. The strength and beauty of DFT is
somewhere else: general considerations about the exact DFT provide us
with a motivation to search for a suitable and physically justified
modelisation $\tilde{E}\left[\rho\right]$ of the exact EDF
$E\left[\rho\right]$. In this way, one is only left with an easy task
of performing the second-stage variation over the density
$\rho(\bm{r})$. Unfortunately, the rigorous link between the exact
many-body Hamiltonian $\hat{H}$ and the model EDF
$\tilde{E}\left[\rho\right]$ is then lost. In this paper, I
propose a method of recovering it.

The goal is thus not to derive the exact EDF $E\left[\rho\right]$, but
to provide an {\abi} derivation valid within a certain class of model EDFs
$\tilde{E}\left[\rho\right]$. Such models should be specific to a given
range of energies or distances, at which low-energy description of ground
states of given physical systems is relevant.

The class I am going to employ is
motivated by 60-odd years of modelling EDFs in nuclei~\cite{[Ben03]}, and can be
formulated as
\be\label{eq:01}
\tilde{E}\left[\rho\right]= \sum_{i=1}^m C^i V_i\left[\rho\right],
\ee
where $C^i$ are coupling constants and $V_i\left[\rho\right]$ are the
Hartree-Fock (or first-order many-body-perturbation-theory) averages
of certain two-body, three-body, etc., operators $\hat{V}_i$. At early
stages of developing the nuclear EDFs, these operators were called
interactions, but in fact, their sole role was to generate specific
terms in the EDF, so here I call them EDF generators.

For the construction presented below, it is essential that the model
EDFs (\ref{eq:01}) are built in terms of true operators acting in the
many-body space, because one must be able to use them not only for
defining the EDFs, but also within the true {\abi} many-body context.
On the one hand, some constructs typical in nuclear EDFs, like the
explicit density-dependent terms~\cite{[Ben03]}, are thus excluded.
On the other hand, functionals based on EDF generators seem to be the
only ones that allow for using EDFs in the multi-reference context,
see, e.g., recent Ref.~\cite{[Bal14a]}, and, therefore, constructions
based on EDF generators are very much called for. We note here that
the proposed scheme would also work for EDFs generated by operators
depending on additional parameters, so the specific linear dependence
on the coupling constants although convenient, is not really
essential.

Before considering specific EDF generators $\hat{V}_i$ that were used
and/or proposed in nuclear physics, let us discuss the main
consequences of using the model EDF in the form of Eq.~(\ref{eq:01}).
First of all, one should keep in mind that the EDF is always meant to
be minimized with respect to the density, and thus its detailed form
beyond the minimum is not essential. By the same token, there is
always a one-to-one correspondence between the coupling constants of
the functional $C^i$ and densities that minimize it. Therefore, the
manifold of meaningful ground-state densities $M[\rho]$ is not really
infinite dimensional, but it can be parametrized by the coupling
constants $C^i$, and eventually by conserved quantum numbers, so it
has a finite number of dimensions. Conversely, the model EDF
(\ref{eq:01}) does not have to properly describe the exact energies
of states having all possible densities, but only those that have
densities on this restricted finite-dimensional manifold $M[\rho]$.

This important observation has far reaching consequences. Indeed,
instead of probing the system with all possible one-body potentials
$-U(\bm{r})$ of an arbitrary shape, as in Eq.~(\ref{eq:00}), it is
enough to probe it within the finite set of the EDF generators
$-\hat{V}_j$, that is, to solve the constrained variational equation,
\be\label{eq:03}
\delta{E'}=\delta\langle\Psi|\hat{H}-\sum_{j=1}^m \lambda^j\hat{V}_j|\Psi\rangle=0,
\ee
for a suitable set of values of a finite number of Lagrange
multipliers $\lambda^i$, which is perfectly manageable a task. In
Eq.~(\ref{eq:03}), there appear the same EDF generators, which in
Eq.~(\ref{eq:01}) were used to define the model EDF in the first
place. This is perfectly logical: to meaningfully include a term in
the model EDF we must first test its properties in the real world of
the {\abi} phase space and Hamiltonian.

Solution of Eq.~(\ref{eq:03}) gives us the exact ground-state
energies $E(\lambda^j)$ and one-body non-local densities
$\rho_{\lambda^j}(\bm{r}_1,\bm{r}_2)$, both as functions (not
functionals!) of the Lagrange multipliers $\lambda^j$. Of course, now
the dependence of densities on Lagrange multipliers cannot be
inverted, however, this is not at all necessary. It is enough to
ensure that, on the manifold generated by the Lagrange multipliers
$\lambda^j$, the model EDF (\ref{eq:01}) best reproduces the exact
energies, that is, it is enough to adjust the EDF coupling constants
$C^i$ so as to have,
\be\label{eq:04}
E(\lambda^j)= \sum_{i=1}^m C^i V_i\left[\rho_{\lambda^j}\right] .
\ee
The adjustment is performed for a finite set of values of the finite
set of Lagrange multipliers, so Eq.~(\ref{eq:04}) constitutes, in
fact, a basic standard linear-regression problem. After the
adjustment, one obtains a true {\abi}-equivalent EDF.

The {\abi} derivation of the model EDFs, proposed in this work, may
become a basis for future studies that can bridge the {\abi} methods
with those related to deriving and improving the phenomenological
EDFs. The proposed research program will probably take some time,
especially in view of the fact that present-day successful {\abi}
implementations are at the forefront of what is currently possible
within the high performance computing. Therefore, in this work I only
present a simple proof-of-principle application of the proposed
scheme to a task of relating one class of the EDF to another.

To this end, I used the EDF generators corresponding to the
central and tensor parts of the nuclear Skyrme
interaction~\cite{[Sky56],[Sky59],[Vau72],[Ben03],[Per04]},
which is composed of eight terms ($m$=8), that is,
\newlength{\dista}
\setlength{\dista}{1.5mm}
\be\label{eq:05}
\left(\!\!\ba{l}
\hat{V}^\rho_0 \\
\hat{V}^\rho_1 \\
\hat{V}^{\Delta\rho}_0 \\
\hat{V}^{\Delta\rho}_1 \\
\hat{V}^\tau_0 \\
\hat{V}^\tau_1 \\
\hat{V}^J_0 \\
\hat{V}^J_1 \\
\ea\!\!\!\right) = \frac{2}{3}
\left(\ba{r@{\hspace{\dista}}r@{\hspace{\dista}}r@{\hspace{\dista}}r@{\hspace{\dista}}r@{\hspace{\dista}}r@{\hspace{\dista}}r@{\hspace{\dista}}r}
 4 & -2  &  0  &  0   &  0  &  0  &  0  &  0 \\
 0 & -6  &  0  &  0   &  0  &  0  &  0  &  0 \\
 0 &  0  & -8  &  4   &  8  & -4  &   8 & -8 \\
 0 &  0  &  0  & 12   &-16  & 20  & -24 & -8 \\
 0 &  0  &  2  & -1   &  6  & -3  &   4 &  0 \\
 0 &  0  &  0  & -3   &-12  & 15  & -12 &  0 \\
 0 &  0  &  0  &  0   &  0  &  0  &  -6 & -6 \\
 0 &  0  &  0  &  0   &  0  &  0  &  18 & -6 \\
\ea\right)\!\!\!
\left(\!\ba{l}
\hat{T}_0         \\
\hat{T}_0^\sigma  \\
\hat{T}_1         \\
\hat{T}_1^\sigma  \\
\hat{T}_2         \\
\hat{T}_2^\sigma  \\
\hat{T}_e         \\
\hat{T}_o         \\
\ea\!\!\!\right),
\ee
where
$\hat{T}_0=\hat{\delta}$,
$\hat{T}_1=\tfrac{1}{2}(\bm{k}'^*{}^2+  \bm{k}^2)\hat{\delta}$,
$\hat{T}_2=(\bm{k}'^*\cdot\bm{k})\hat{\delta}$,
and $\hat{\delta}$ is a two-body local zero-range potential,
$\hat{\delta}=\delta(\bm{r}_1-\bm{r}_2)\delta(\bm{r}_1-\bm{r}'_1)\delta(\bm{r}_2-\bm{r}'_2)$.
The standard relative-momentum operators are defined as $\bm{k}=(\bm{\nabla}_1-\bm{\nabla}_2)/2i$
and $\bm{k}'=(\bm{\nabla}'_1-\bm{\nabla}'_2)/2i$, $\hat{T}_i^\sigma=\hat{T}_i\hat{P}_{\sigma}$ for
$\hat{P}_{\sigma}=\tfrac{1}{2}(1+\bm{\sigma}_1\cdot\bm{\sigma}_2)$, and
$\hat{T}_e =\tfrac{1}{2}
({\bm{k}}^{\prime *}\cdot\hat{{\mathsf S}}\cdot{\bm{k}}^{\prime *}
+  \bm{k} \cdot\hat{{\mathsf S}}\cdot\bm{k})$ and
$\hat{T}_o ={\bm{k}}^{\prime *}\cdot\hat{{\mathsf S}} \cdot\bm{k}$ for
$\hat{{\mathsf S}}^{ab}
=\tfrac{3}{2}\big(\sigma_1^a
                  \sigma_2^b
                 +\sigma_1^b
                  \sigma_2^a\big)
-  \delta_{ab}\bm{\sigma}_1\cdot
              \bm{\sigma}_2$.

Numerical coefficients appearing in Eq.~(\ref{eq:05}) were chosen in
such a way that each of the eight EDF generators gives (in spherical
nuclei) one specific term of the EDF~\cite{[Eng75],[Bei75],[Per04]},
namely,
\be\label{eq:06}\ba{rclcrcl}
{V}^\rho_t\left[\rho\right]&=&\rho_t(\bm{r})\rho_t(\bm{r}) &,\quad&
{V}^{\Delta\rho}_t\left[\rho\right]&=&\rho_t(\bm{r})\Delta\rho_t(\bm{r}), \\
{V}^\tau_t\left[\rho\right]&=&\rho_t(\bm{r})\tau_t(\bm{r}) &,\quad&
{V}^{J}_t\left[\rho\right]&=&{\mathsf J}_t(\bm{r})\cdot{\mathsf J}_t(\bm{r}), \\
\ea\ee
where index $t$ refers to the isoscalar ($t$=0) or isovector ($t$=1) densities, $\Delta\rho_t(\bm{r})$
stands for the Laplacian of the density, and ${\tau}_t(\bm{r})=\big[(\bm{\nabla}\cdot\bm{\nabla}')
{\rho}_t(\bm{r},\bm{r}')\big]_{\bm{r}=\bm{r}'}$ and
${{\mathsf J}}^{ab}_t(\bm{r})=\tfrac{1}{2i}\big[ (\nabla_a - \nabla_a')
s^b_t(\bm{r},\bm{r}')\big]_{\bm{r}=\bm{r}'}$
are the standard quasilocal kinetic and spin-current densities, respectively.

For the proof-of-principle application presented
in this work, instead of the average value of the true many-body
Hamiltonian, I used the Gogny EDF~\cite{[Gog75]} in the D1S
parametrization~\cite{[Ber91b]}, that is,
$\langle\Psi|\hat{H}|\Psi\rangle\rightarrow{}E_{\text{Gogny}}\left[\rho\right]$.
In this way, I aimed at obtaining a Gogny-equivalent quasilocal
Skyrme EDF. That both types of EDFs can be linked to one another is
already known from findings of Refs.~\cite{[Car10e],[Dob12a]}, where
this fact was demonstrated within the DME and effective theory; here I aim
at testing this equivalence in terms of the {\abi}-like methodology
proposed in this work. Since both functionals contain terms generated
by the zero-range spin-orbit and density-dependent operators, these
were left untouched, and in the left-hand side of Eq.~(\ref{eq:04})
only the part of the Gogny EDF generated by the finite-range
potentials was used.

Numerical results presented below were obtained using the code
{\sc hfodd} (v2.75c)~\cite{[Sch15]}, which is the only existing code
capable of treating the Gogny and Skyrme functionals simultaneously
and within the same numerical infrastructure, see the Supplemental
Material~\cite{suppl-abi} for details. Calculations were performed for eight doubly magic nuclei,
$^{16}$O, $^{40,48}$Ca, $^{56,78}$Ni, $^{100,132}$Sn,
and  $^{208}$Pb.
For each nucleus, I used
either of the eight Lagrange multipliers,
${\lambda}_t^\rho$,
${\lambda}_t^{\Delta\rho}$,
${\lambda}_t^\tau$,  or
${\lambda}_t^J$ (for $t$=0,1) equal to one of the 21 integer values between $-$10 and +10\,MeV\,fm$^n$, where $n$=3 for
${\lambda}_t^\rho$ and $n$=5 for the other ones.
Altogether, this gave me 1344 values
of the Gogny energies $E(\lambda^j)$, to which the eight coupling constants
of the Skyrme EDF,
$C_t^\rho$,
$C_t^{\Delta\rho}$,
$C_t^\tau$, and
$C_t^J$       (for $t$=0,1) were adjusted in Eq.~(\ref{eq:04}).

\begin{table}
\begin{center}
\caption{\label{tab3}
The Skyrme EDF S1Sd and S1Se coupling constants obtained in this work as
the {\abi}-equivalent Gogny EDF D1S~\protect\cite{[Ber91b]}.
}
\vspace*{1ex}
\begin{tabular}{lrrrcrr}
\hline
         &&  \multicolumn{2}{c}{S1Sd} &~& \multicolumn{2}{c}{S1Se}  \\
\cline{3-4} \cline{6-7}
         &&  \multicolumn{1}{c}{$t=0$} & \multicolumn{1}{c}{$t=1$}
         &&  \multicolumn{1}{c}{$t=0$} & \multicolumn{1}{c}{$t=1$}  \\
$C^\rho_t$        & (MeV\,fm$^3$)& $-$603.82(22)   &  $ $484(4)&& $-$605.41(16)     &  $ $509(3) \\
$C^{\Delta\rho}_t$& (MeV\,fm$^5$)& $-$73.25(14)    &  $ $48(3) && $-$74.82(12)      &  $ $41(2)  \\
$C^\tau_t$        & (MeV\,fm$^5$)& $ $77.95(23)    &  $-$78(3) && $ $79.73(16)      &  $-$98(2)  \\
$C^J_t$           & (MeV\,fm$^5$)& $ $24.9(1.2)    &  $ $71(3) && $ $ 0             &  $ $ 0     \\
\hline
\end{tabular}
\end{center}
\end{table}

\begin{figure}
\includegraphics[width=\columnwidth]{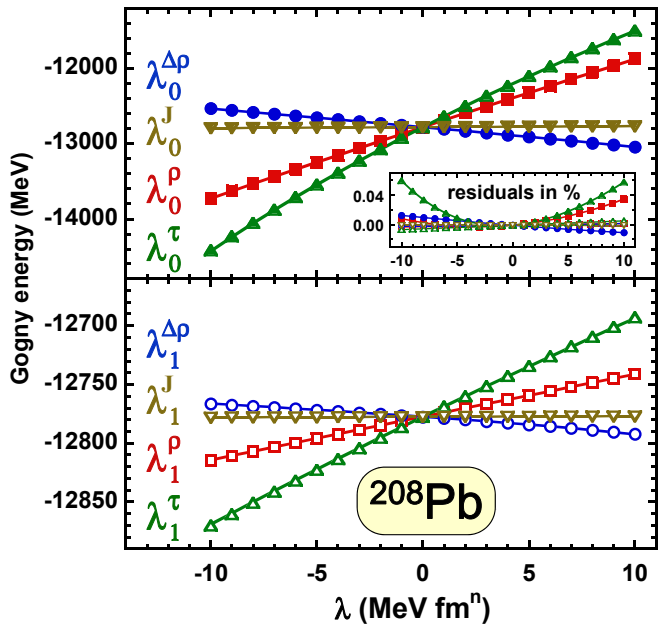}
\caption{(Color online) Gogny energies [lines, left-hand side of
Eq.~(\protect\ref{eq:04})] compared with the EDF estimates [symbols,
right-hand side of Eq.~(\protect\ref{eq:04})] obtained for the Skyrme
EDF S1Sd coupling constants given in Table~\protect\ref{tab3}.
Calculations were performed in $^{208}$Pb in function of the eight
Lagrange multipliers
${\lambda}_t^\rho$,
${\lambda}_t^{\Delta\rho}$,
${\lambda}_t^\tau$, and
${\lambda}_t^J$ for $t$=0,1. The inset shows residuals of the adjustment in per cent.
}
\label{fig:1}
\end{figure}

The coupling constants S1Sd obtained by such an adjustment are shown
in Table~\ref{tab3}. Their standard uncertainties were obtained within the
standard regression analysis presented, e.g., in Refs.~\cite{[Toi08],[Dob14a]}.
The adjusted coupling constants reproduced the Gogny energies
in Eq.~(\ref{eq:04}) with a very high accuracy: the relative rms deviation
between left- and right-hand sides of Eq.~(\ref{eq:04}) is only 0.014\%.
This is also illustrated in Fig.~\ref{fig:1}, where differences between
symbols and lines cannot be seen at all, while the inset shows that
the relative residuals of the adjustment do not exceed 0.050\%.
Analogous plots for other nuclei are collected in the Supplemental
Material~\cite{suppl-abi}.

\begin{table}
\begin{center}
\caption{\label{tab4a}
Gogny EDF D1S ground-state energies $E_G$ (b) of eight doubly magic nuclei (a)
compared to energies $E$ (c) calculated
using the Skyrme EDF S1Sd, Table~\protect\ref{tab3}, and shown together
with their propagated uncertainties $\Delta{E}$.
Column (d) shows the residuals $\delta{E}=E-E_G$ and
columns (e) and (f) give ratios of residuals
with respect to energies $E$ and propagated uncertainties $\Delta{E}$,
respectively. All energies are in MeV.
}
\vspace*{1ex}
\begin{tabular}{lrcrrrr}
\hline
             \multicolumn{1}{l}{}
          &  \multicolumn{1}{c}{$E_G$}
          && \multicolumn{1}{c}{$E$}
          &  \multicolumn{1}{c}{$\delta{E}$}
          &  \multicolumn{1}{c}{$\delta{E}/|E|$}
          &  \multicolumn{1}{c}{$\delta{E}/\Delta{E}$} \\
             \multicolumn{1}{l}{(a)}
          &  \multicolumn{1}{c}{(b)}
          && \multicolumn{1}{c}{(c)}
          &  \multicolumn{1}{c}{(d)}
          &  \multicolumn{1}{c}{(e)}
          &  \multicolumn{1}{c}{(f)} \\
\hline
$^{ 16}$O  & $-$129.626   &&  $-$129.56(4)~ & $ $0.07 & $ $0.05\% &$ $ 2  \\
$^{ 40}$Ca & $-$344.663   &&  $-$346.01(6)~ & $-$1.35 & $-$0.39\% &$-$23  \\
$^{ 48}$Ca & $-$416.829   &&  $-$418.10(6)~ & $-$1.27 & $-$0.30\% &$-$20  \\
$^{ 56}$Ni & $-$483.820   &&  $-$485.31(6)~ & $-$1.49 & $-$0.31\% &$-$27  \\
$^{ 78}$Ni & $-$640.598   &&  $-$642.37(6)~ & $-$1.78 & $-$0.28\% &$-$28  \\
$^{100}$Sn & $-$830.896   &&  $-$833.19(6)~ & $-$2.29 & $-$0.28\% &$-$39  \\
$^{132}$Sn & $-$1103.246  &&  $-$1106.51(8) & $-$3.26 & $-$0.30\% &$-$42  \\
$^{208}$Pb & $-$1638.330  &&  $-$1640.96(8) & $-$2.63 & $-$0.16\% &$-$32  \\ \hline
rms        &    n.a.      &&     n.a.       & $ $1.99 & $ $0.28\% &$ $29  \\
\hline
\end{tabular}
\end{center}
\end{table}

Table~\ref{tab4a} compares the ground-state energies $E_G$ calculated
using the original Gogny EDF D1S with energies $E$ obtained by the
minimization of the Gogny-equivalent Skyrme EDF S1Sd. Propagated
uncertainties $\Delta{E}$ of $E$ were calculated using the covariance matrix
related to the adjustment of coupling
constants~\cite{[Toi08],[Dob14a]}, see the Supplemental
Material~\cite{suppl-abi}. We see that the Skyrme EDF S1Sd again
reproduces the Gogny-EDF results with a very high accuracy: the
relative rms deviations between these two functionals is only 0.28\%.
This is much better than the accuracy obtained within the
DME~\cite{[Dob10]}, see the comparison presented in the Supplemental
Material~\cite{suppl-abi}. One can conclude that the simple
eight-dimensional Gogny-equivalent Skyrme EDF with {\abi}-derived
coupling constants of Eq.~(\ref{eq:01}) very well describes the full
Gogny energies. We note here that the coupling constants were
adjusted only to energies. In the Supplemental
Material~\cite{suppl-abi}, I also show the analogous very good
agreement obtained for the proton rms radii. This points to a
well-built EDF, which along with the total energies properly
describes one-body observables.

On the absolute scale, the corresponding rms deviation of energies is
1.99\,MeV, which is 29 times higher than the rms average of the
propagated uncertainties, see Table~\ref{tab4a}. It means that the
differences between the Gogny results and Gogny-equivalent Skyrme-EDF
results are still significantly larger than the uncertainties of
the adjustment, which gives a clear signal for missing terms in the
eight-dimensional model EDF of Eq.~(\ref{eq:04}).

The method also allows for testing the impact of removing terms from
the model EDF. For example, setting the two spin-current coupling
constants to zero, $C_0^J=C_1^J=0$, one obtains a six-dimensional
model that gives another Gogny-equivalent Skyrme EDF S1Se, with
coupling constants shown in Tables~\ref{tab3}. Such model is only
marginally worse, with the absolute and relative rms deviations of
energies now increased to 2.34\,MeV and 0.40\%, respectively, see the
Supplemental Material~\cite{suppl-abi} for detailed results.

In conclusion, I proposed a novel method of obtaining
{\abi}-equivalent model EDFs. The main idea is in replacing the
standard-DFT use of an external one-body potential by the use of
two-body, three-body, etc., EDF generators. This probes the
reaction of the system with respect to the same operators that are
used to construct the model EDFs. The new method
amounts to performing {\abi} calculations with simple constraints on
a finite set of well-defined operators added to the many-body
Hamiltonian, and thus is perfectly manageable.

The method is also able to give us a quantitative information on
whether a given model EDF is adequate for the proper description of
the physical system being studied. Indeed, if the adjustment of the
coupling constants fails to be accurate enough, we obtain a clear
signal that the set of proposed EDF generators is inadequate. Then,
another or extended model EDF should be tried. In this way, through
{\abi} derivations in nuclei, one may be able to evaluate relative
merits of using the zero-range higher-order~\cite{[Rai11b]},
finite-range regularized~\cite{[Rai14]}, three-body or
four-body~\cite{[Sad13a]}, or gradient-dependent
three-body~\cite{[Sad13b]} pseudopotentials, which are currently
being developed and implemented in nuclear EDF approaches.

It is very important that the proposed method is based on studying
specific finite systems and does not rely on assumptions valid only
in infinite or semi-infinite systems. The proposed {\abi} derivations
can be performed in few systems, for which the {\abi} calculations
are possible, e.g., in light closed-shell nuclei, and then the
derived EDFs can be applied to more complicated, open-shell or heavy
nuclei, so as to test the overall predictive power of the method. It
is also essential that the {\abi}-equivalent EDFs are specific to
particular physical systems, and when applied to systems at
different energies or densities will yield parametrically energy- or
density-dependent running coupling constants. Needless to say that
the proposed method can easily be extended to deriving time-odd or
pairing terms of the EDFs. Another fascinating extension would be
to use the same method not only to match energies, like in Eq.~(\ref{eq:04}),
but also {\abi}-derived kernels~\cite{[Dug15a],[Dug15b]}.

\begin{acknowledgments}
Very interesting discussions with Witek Nazarewicz are gratefully acknowledged.
This work was supported in part
by the Academy of Finland and University of Jyv\"askyl\"a within the FIDIPRO program,
by the Polish National Science Center under Contract No.\ 2012/07/B/ST2/03907, and
by the ERANET-NuPNET grant SARFEN of the Polish National Centre for Research and Development (NCBiR).
We acknowledge the CSC-IT Center for Science Ltd., Finland, for the allocation of
computational resources.
\end{acknowledgments}


\end{document}